\documentclass[12pt]{article}    % LaTeX 2e
\usepackage{latexsym} % Gets \Box etc
\usepackage{amssymb}  % \Box etc: see /usr/lib/texmf/tex/ams/doc/amsguide.ps
\usepackage{amsfonts} % \mathfrak and \mathbb
\usepackage{multirow}
\usepackage{epsf}

\newcommand{\al}{\alpha}
\newcommand{\be}{\beta}
\newcommand{\ga}{\gamma}

\newcommand{\de}{\delta}
\newcommand{\De}{\Delta}

\newcommand{\ka}{\kappa}
\newcommand{\la}{\lambda}
\newcommand{\La}{\Lambda}

\newcommand{\si}{\sigma}
\newcommand{\Si}{\Sigma}
\newcommand{\<}{\langle} 
\renewcommand{\>}{\rangle} % LaTeX: \> already defined

\newcommand{\txt}{\textstyle}
\newcommand{\dsp}{\displaystyle}

\newcommand\eqn[1]{(\ref{#1})}      % parentheses around the LaTex "ref" macro
\newcommand\Eqn[1]{Eq.~(\ref{#1})}  % includes ``Eq.'' in front

\newcommand{\beq}{\begin{equation}}
\newcommand{\eeq}{\end{equation}}
\newcommand{\ba}{\begin{array}}
\newcommand{\bea}{\begin{eqnarray}}
\newcommand{\ea}{\end{array}}
\newcommand{\eea}{\end{eqnarray}}

\newcommand\comment[1]{ \hbox{[{\it Comment suppressed here.}\/]} }
\newcommand\hide[1]{}

\newcommand{\skipover}[1]{}

\newcommand{\half} {{\txt {1\over 2}}}

\newcommand{\MeV}{{\rm MeV}}
\newcommand{\GeV}{{\rm GeV}}

\def\appendix{\par                              % Have \appendix say
    \setcounter{section}{0}                     % `Appendix A', not just `A'
    \setcounter{subsection}{0}
    \renewcommand{\theequation}{\Alph{section}.\arabic{equation}}
    \renewcommand{\thesection}{Appendix \Alph{section}\setcounter{equation}{0}}
}

\catcode`\@=11

\def\applabel#1{\@bsphack
  \protected@write\@auxout{}%
         {\string\newlabel{#1}{{\Alph{section}}{\thepage}}}%
  \@esphack}

\def\section{
\setcounter{equation}{0}        % Reset eqn numbers at start of section
\@startsection {section}{1}{\z@}{-3.5ex plus -1ex minus 
 -.2ex}{2.3ex plus .2ex}{\large\bf}}
\renewcommand{\theequation}{\arabic{section}.\arabic{equation}}

\def\subsection{\@startsection{subsection}{2}{\z@}{-3.25ex plus -1ex minus 
 -.2ex}{1.5ex plus .2ex}{\normalsize\bf}}

\def\subsubsection{\@startsection{subsubsection}{3}{\z@}{-3.25ex plus
 -1ex minus -.2ex}{1.5ex plus .2ex}{\normalsize}}

\newsavebox{\eqlabel}
\makeatletter  %\catcode`\@=11
\newlength{\numblen}
\newsavebox{\eqnumb}
\def\@eqnnum{\savebox{\eqnumb}{\rm (\theequation)}%
\settowidth{\numblen}{\usebox{\eqnumb}}%
\makebox[\numblen][l]{\usebox{\eqnumb}~~~\usebox{\eqlabel}}}
\makeatother   %\catcode`\@=12

\newenvironment{equationwithlabel}[1]{ %
  \begin{equation}\label{#1} }{\end{equation}} %\savebox{\eqlabel}{~}}
\newcommand{\beql}[1]{\begin{equationwithlabel}{#1}}
\newcommand{\eeql}{\end{equationwithlabel}}
\newcommand{\ms}{m_s}
\newcommand{\Ms}{M_s}
\newcommand{\Minv}{{M^{-1}}}

\newcommand{\qplslash}{{q\!\!\!/}_+}
\newcommand{\qmislash}{{q\!\!\!/}_-}
\newcommand{\qvsq}{{\vec q}^{\,2}}

\begin{document}

\title{\bf Unlocking Color and Flavor\\ 
in\\ Superconducting Strange Quark Matter}

\newcommand{\ns}{\normalsize}

\author{
Mark Alford, J\"urgen Berges, Krishna Rajagopal \\[0.5ex]
{\normalsize Center for Theoretical Physics,}\\
{\normalsize Massachusetts Institute of Technology, Cambridge, MA 02139 }
}

\newcommand{\preprintno}{\normalsize 
MIT-CTP-2844
}

\date{\preprintno}

\begin{titlepage}
\maketitle
\def\thepage{}          % No page number on title page

\begin{abstract}
%       1         2         3         4         5         6

We explore the phase diagram of strongly interacting matter with
massless $u$ and $d$ quarks as a function of the strange quark mass
$m_s$ and the chemical potential $\mu$ for baryon number.  Neglecting
electromagnetism, we describe the different baryonic and quark matter
phases at zero temperature. For quark matter, we support our
model-independent arguments with a quantitative analysis of a model
which uses a four-fermion interaction abstracted from single-gluon
exchange.  For any finite $m_s$, at sufficiently large $\mu$ we find
quark matter in a color-flavor locked state which leaves a global
vector-like $SU(2)_{{\rm color}+L+R}$ symmetry unbroken. As a
consequence, chiral symmetry is always broken in sufficiently dense
quark matter.  As the density is reduced, for sufficiently large $m_s$
we observe a first order transition from the color-flavor locked phase
to a color superconducting phase analogous to that in two flavor QCD.
At this unlocking transition chiral symmetry is restored.  For
realistic values of $m_s$ our analysis indicates that chiral symmetry
breaking may be present for all densities down to those characteristic
of baryonic matter.  This supports the idea that quark matter and
baryonic matter may be continuously connected in nature. We map
the gaps at the quark Fermi surfaces in the high density color-flavor
locked phase onto gaps at the baryon Fermi surfaces at
low densities.

\end{abstract}

\end{titlepage}

\renewcommand{\thepage}{\arabic{page}}
%\setcounter{page}{1}

%--------------------------------------- 
                        % Body of paper begins

\section{Introduction and Phase Diagram}
\label{sec:int}
Strongly interacting matter at high baryon number density and low
temperature is far less well understood than strongly interacting
matter at high temperature and zero baryon number density.  At high
temperatures, the symmetry of the lowest free energy state is not in
dispute, calculations using lattice gauge theory are bringing the
equilibrium thermodynamics under reasonable quantitative control, and
even non-equilibrium dynamical questions are being addressed. 
At high densities, we are still learning about the symmetries
of the lowest free energy state.   In addition to being of
fundamental interest, an understanding of the symmetry properties
of dense matter can be expected to inform our understanding
of neutron star astrophysics and perhaps also heavy
ion collisions which achieve high baryon densities without
reaching very high temperatures.

At high densities and low
temperatures, the relevant degrees of freedom are those which involve
quarks with momenta near the Fermi surface(s).  The presence of an
arbitrarily weak attraction between pairs of quarks results in the
formation of a condensate of quark Cooper pairs. Pairs of quarks cannot
be color singlets, and in QCD with two
flavors of massless quarks the Cooper pairs form in the
color ${\bf \bar 3}$ 
channel.\cite{Barrois,BailinLove,ARW2,RappETC,BergesRajagopal}
The resulting condensate gives gaps to quarks with two of three
colors and
breaks
the local color symmetry 
$SU(3)_{\rm color}$ to an $SU(2)_{\rm color}$ 
subgroup.
The breaking of a gauge symmetry
cannot be characterized by a gauge invariant local
order parameter which vanishes on one side of
a phase boundary. The superconducting phase can 
be characterized rigorously only by its global symmetries.
In QCD with two flavors,  the Cooper pairs are $ud-du$ flavor
singlets and, in particular, the global flavor
$SU(2)_L \times SU(2)_R$ 
symmetry is left intact.
There is an
unbroken global symmetry which plays the role of baryon number
symmetry, $U(1)_B$.
Thus, no global symmetries are broken and the
only putative Goldstone bosons are those five which become the
longitudinal parts of the five gluons which 
acquire masses.\cite{ARW2}\footnote{There is
also an unbroken gauged symmetry which plays the role of electromagnetism.
Also, the third color quarks can condense,\cite{ARW2} as we
discuss below, but the resulting gap is much smaller.}

In QCD with three flavors of massless quarks 
the Cooper pairs {\it cannot}
be flavor singlets, and both color and flavor symmetries are
necessarily broken.   The symmetries of the phase which
results have been analyzed in Ref.~\cite{ARW3}.
The attractive channel favored by one-gluon
exchange exhibits ``color-flavor locking''.
It locks $SU(3)_L$ flavor rotations to $SU(3)_{\rm color}$,
in the sense that the condensate is not symmetric under
either alone, but is symmetric under the 
simultaneous $SU(3)_{{\rm color}+L}$ rotations.
The condensate
also locks $SU(3)_R$ rotations to $SU(3)_{\rm color}$,
and since color is a vector symmetry 
the chiral $SU(3)_{L-R}$ symmetry is broken.
Thus, in quark matter with three massless quarks,  the 
$SU(3)_{\rm color}\times SU(3)_{L}\times
SU(3)_{R}\times U(1)_B$ symmetry is broken down to the global
diagonal $SU(3)_{{\rm color}+V}$ subgroup.  
There is also an unbroken gauged $U(1)$ symmetry (under which
all quarks have integer charges) which plays the role
of electromagnetism.
All nine quarks have a gap.
All eight gluons get a mass. There are nine
massless Nambu Goldstone excitations of the condensate of Cooper pairs
which result from the breaking of the axial $SU(3)_A$ and 
baryon number $U(1)_B$.  
We see that cold
dense quark matter has rather different 
global symmetries for $m_s=0$ than
for $m_s=\infty$.  

A nonzero strange quark mass explicitly breaks the 
flavor $SU(3)_V$ symmetry. As a consequence, color-flavor locking
with an unbroken global $SU(3)_{{\rm color}+V}$ occurs only for
$m_s\equiv 0$. Instead, for nonzero but sufficiently small strange 
quark mass we expect, and find, color-flavor locking  
which leaves a global $SU(2)_{{\rm color}+V}$ group unbroken.
As $m_s$ is increased from zero to infinity, there has to be some 
value $m_s^{\rm unlock}$ at which color and flavor rotations 
are unlocked, and 
the full $SU(2)_L \times SU(2)_R$ symmetry is restored.
We argue on general grounds in Section \ref{sec:general} that
this unlocking phase transition must be of first order.
In subsequent Sections, we
analyze this transition %in sections \ref{sec:con}-\ref{sec:sol}
quantitatively in a model using a four-fermion interaction 
with quantum numbers abstracted from single-gluon exchange.

{}From our analysis of the unlocking transition, we conclude that
for realistic values of the strange quark mass 
chiral symmetry breaking may be present for
densities all the way down to those characteristic of baryonic matter.
This raises the possibility that
quark matter and baryonic matter may be continuously 
connected in nature, as Sch\"afer and Wilczek 
have conjectured 
for QCD with three massless quarks.\cite{SchaeferWilczek}
We use our calculations of the properties of 
color-flavor locked superconducting
strange quark matter to map the gaps due to pairing at the quark
Fermi surfaces onto gaps
due to pairing at the baryon Fermi surfaces in 
superfluid baryonic matter consisting of nucleons, 
$\Lambda$'s, $\Sigma$'s, 
and $\Xi$'s. (See Section~\ref{sec:cont}).

We argue that  color-flavor 
locking will always occur 
for sufficiently large chemical potential, for 
any nonzero, finite $m_s$.  We make
this argument by first using our results as a guide 
to quark matter at moderate
densities and then using them to normalize 
Son's model-independent analysis valid at 
very high densities.\cite{Son}
As a consequence of color-flavor locking,
chiral symmetry is spontaneously broken even at asymptotically
high densities, in sharp contrast to the well established
restoration of chiral symmetry at high temperature.

\begin{figure}[thb]
\epsfxsize=5in
\begin{center}
\hspace*{0in}
\epsffile{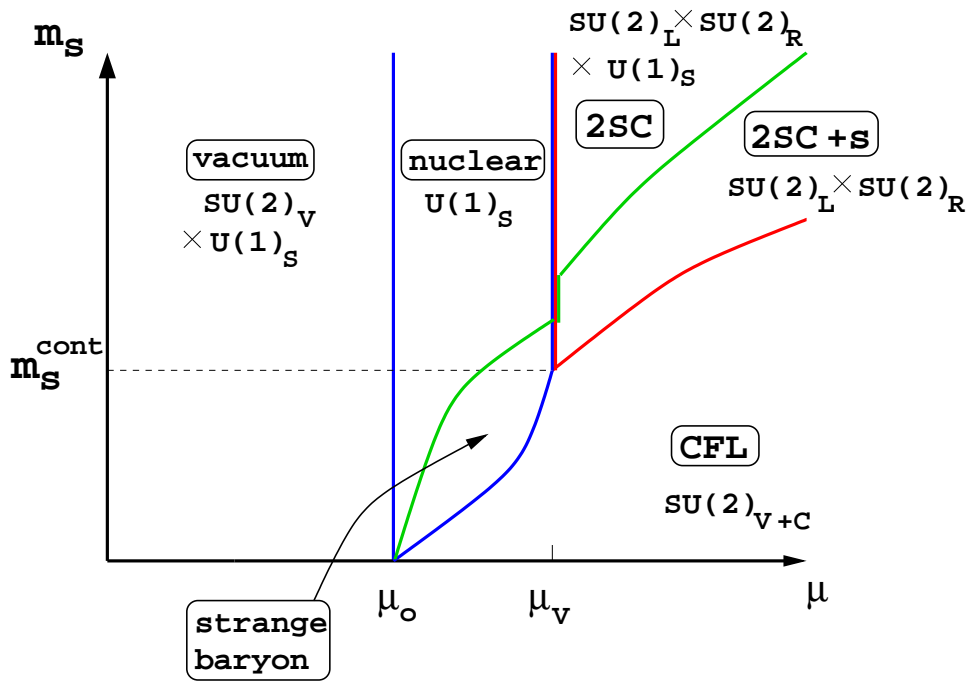}
%\vspace{-2ex}
\end{center}
\caption{
Conjectured phase diagram for QCD with two massless quarks
and a strange quark at zero temperature. 
The global symmetries of each phase are labelled.
The regions of the phase diagram labelled 
2SC, 2SC+s and CFL denote color superconducting quark matter phases.
The Figure is described at length in the text.
} 
\label{fig:phasediagram}
\end{figure}
Our goal in this paper is a consistent
picture of the symmetries of cold dense quark matter with 
massless $u$ and $d$ quarks as a function of
the strange quark mass $m_s$ and the chemical potential $\mu$.
We take $\mu$ to be the chemical potential for quark number,
one third that for baryon number. $m_s$ is the current
quark mass; we will refer to the $\mu$-dependent constituent
quark mass as $M_s(\mu)$.

Figure 1 summarizes our conjecture for the zero temperature
phase diagram of QCD as a function 
of $m_s$ and $\mu$. (We work at zero temperature throughout
this paper.)
The rest of this section can be read as a description
of this Figure. Lines in the diagram separate phases which
differ in their global symmetries. This means that each line
describes a distinction which can be associated with 
a local order parameter which vanishes
on one side of the line. 
In each region of the diagram, we list 
the unbroken global symmetries of the corresponding phase.
We characterize the phases using the $SU(2)_L\times SU(2)_R$ 
flavor rotations of the light quarks, and the 
$U(1)_S$ rotations of the strange quarks.\footnote{The
$U(1)_B$ symmetry associated
with baryon number is a combination of $U(1)_S$, a $U(1)$ subgroup of
isospin, and the gauged $U(1)_{\rm EM}$ of electromagnetism.
Therefore, in our analysis of the global symmetries, 
once we have analyzed isospin and strangeness,
considering baryon number adds nothing new.}

In subsequent Sections of this paper, we 
explore those
phases (labelled 2SC, 2SC+s and CFL)
which extend to high enough density that they
can be described as superconducting quark matter. 
We analyze these phases, and the unlocking phase transition
which separates them, quantitatively in a 
model in which quarks interact via a four-quark interaction
modelled on that induced 
by single-gluon exchange.  Certainly, our model is not
expected to be a valid 
description for QCD at nuclear matter density,
where confinement plays an important role. That part of the
diagram which describes the symmetry properties
of different phases of baryonic matter is
therefore not derived from our model. It is 
conjectural but plausible.

In Figure 1 and throughout this paper, we  
neglect the small $u$ and $d$ current quark masses.
The light quark masses have no
substantial influence on the condensation of quark Cooper 
pairs.\cite{BergesRajagopal,PisarskiRischke1OPT}  We also ignore the
effects of electromagnetism throughout. 
We assume that wherever a baryon Fermi surface is
present, baryons always pair at zero temperature.
To simplify our analysis, we assume that baryons always
pair in channels which preserve rotational invariance,
breaking internal symmetries such as isospin if necessary. 

We now explain the features shown in  Figure 1 
by beginning at $\mu=0$ (in vacuum) and describing the
phase transitions which occur as $\mu$ is increased at constant $m_s$.
We do this twice, first with a large enough value of $m_s$ 
that the strange quark is heavy enough 
that 
immediately above 
deconfinement $\mu$ is still less than $m_s$ and
there are still no strange quarks present.
For $\mu=0$ the density is
zero; isospin and strangeness are unbroken; Lorentz symmetry is
unbroken; chiral symmetry is broken.
Above a first order
transition\footnote{Discussed in Ref.\ \cite{Halasz}.} 
at an onset chemical potential $\mu_{\rm o}\sim 300~\MeV$, 
one finds nuclear matter (``nuclear'' in Figure~\ref{fig:phasediagram}).  
Lorentz symmetry is broken, leaving only
rotational symmetry manifest. 
Chiral symmetry is thought to be 
broken, although the
chiral condensate $\langle \bar q q \rangle$ is expected to be reduced
from its vacuum value.
In the nuclear matter phase, we expect an instability of the nucleon
Fermi surfaces to lead to Cooper pairing. We assume that
(as is observed in nuclei) the pairing is 
$pp$ and $nn$, breaking
isospin.  
Since there are no 
strange baryons present, $U(1)_S$
is unbroken.

At the large value of $m_s$ we are currently describing, 
when $\mu$ is increased above $\mu_{\rm V}$, we 
find the ``2SC'' phase  of color-superconducting
matter consisting of up and down quarks only, described
in Refs.~\cite{Barrois,BailinLove,ARW2,RappETC}.
A nucleon description is no longer appropriate. 
The light quarks pair in isosinglet channels. $SU(2)_V$
is unbroken. $SU(2)_A$ is unbroken. 
Quarks of two colors
pair in a Lorentz singlet channel; quarks carrying the
third color 
can form an axial vector (ferromagnetic) condensate, which
breaks rotational invariance.  The associated gap
is of order a keV or much less\cite{ARW2} and we neglect
this condensate in Figure 1.\footnote{In a mean-field
treatment, one gluon exchange
is neither attractive nor repulsive in the channel
which leads to the ferromagnetic condensate while the instanton
interaction is only weakly attractive. The 
ferromagnetic condensate is therefore fragile in the sense
that this channel could easily be rendered repulsive
by other interactions.\cite{ARW2} If that were the case,
the third color quarks would presumably form Cooper pairs
with nonzero orbital angular momentum, and an even smaller gap.}  
The phase transition at $\mu_V$ is first 
order
\cite{ARW2,RappETC,BergesRajagopal,BJW2,PisarskiRischke1OPT,CarterDiakonov}
and is characterized by a competition between the chiral
$\langle \bar q q\rangle$ condensate and the superconducting
$\langle q q \rangle$ condensate.\cite{BergesRajagopal,CarterDiakonov}

As the chemical potential is increased further, when $\mu$
exceeds the constituent strange quark mass $M_s(\mu)$
a strange quark Fermi surface forms, with a Fermi momentum
far below that for the light quarks.  We denote the resulting
phase ``2SC+s''.
Light and strange quarks do not pair with each other, because
their respective Fermi momenta are so different 
(see Section \ref{sec:general}).
The strange Fermi
surface is presumably nevertheless unstable.  The resulting
$ss$ condensate must be constructed from Cooper pairs which are
either color ${\bf 6}$, or have spin $1$, or have nonzero 
angular momentum, or must be $\langle s C \gamma_4 s \rangle$,
which is symmetric in Dirac indices.
Each of these options lead to small gaps, for different
reasons: One-gluon exchange
is repulsive in rotationally invariant 
color ${\bf 6}$ channels and the $\langle s C \gamma_4 s \rangle$
channel, and although
other interactions may overcome this and result in a net
attraction, this is likely to be much weaker than the
attraction in the dominant color ${\bf \bar 3}$ channels. Gaps
involving Cooper pairs with nonzero $J$ tend to be 
significantly suppressed because not all quarks at the Fermi surface
participate.\cite{ARW2} 
Thus, although $U(1)_S$ will be broken
in the ``2SC+s'' phase by an $ss$ condensate in {\it some}
channel, we expect that the resulting gaps will be very small.
In our analysis below we therefore
neglect the difference between the 2SC and 2SC+s 
phases.

Finally, when the chemical potential is high enough
that the Fermi momenta for the strange and light quarks
become comparable, we cross the first order locking
transition described in detail
in the next  Sections, and find 
the color-flavor locked (CFL) phase.
There is an unbroken global symmetry constructed
by locking the $SU(2)_V$ isospin rotations and an
$SU(2)$ subgroup of color. Chiral symmetry is once again broken.

We now describe the sequence of phases which arise
as $\mu$ is increased,
this time for a value of $\ms$
small enough that strange baryonic matter
forms below the deconfinement density. 
At $\mu_{\rm o}$, one enters the nuclear matter phase, 
with the familiar
$nn$ and $pp$ pairing at the neutron and proton Fermi surfaces
breaking isospin.  
The $\Lambda$, $\Sigma$ and $\Xi$ densities are
still zero, and strangeness is unbroken.  At a somewhat larger
chemical potential, we enter the strange baryonic matter phase, with
Fermi surfaces for the $\La$ and $\Si$. These pair with themselves in
spin singlets, breaking $U(1)_S$.  This phase is labelled ``strange
baryon'' in Figure 1.  The global symmetries $SU(2)_L\times SU(2)_R$
and $U(1)_S$ are all broken. 
As $\mu$ rises, one
finds yet another onset at which the $\Xi$ density becomes
nonzero. This breaks no new symmetries, and so is not shown in the Figure.
Note that kaon condensation\cite{KaplanNelson} breaks $U(1)_S$, and $SU(2)_V$.
The only phase in the diagram in which this occurs is that which
we have labelled the
strange baryon phase.  Thus, if kaon condensation occurs, by definition it
occurs within this region of the diagram.  
If kaon condensation is favored, this will
tend to enlarge the region of the diagram within which $U(1)_S$ and 
$SU(2)_V$ are both broken.

We can imagine two possibilities for what happens next as $\mu$ increases
further.
(1) Deconfinement: the baryonic Fermi surface is replaced by
$u,d,s$ quark Fermi
surfaces, which are unstable against pairing, and
we enter the CFL phase, described above. Isospin is locked to color and
$SU(2)_{{\rm color}+V}$ is restored, but chiral symmetry remains broken.
(2) No deconfinement: the Fermi momenta of all of the octet
baryons are now similar enough that pairing between baryons with
differing strangeness becomes possible.  At this point,
isospin is restored: the baryons pair in rotationally
invariant isosinglets
($p\Xi^-$, $n\Xi^0$, $\Si^+ \Si^-$, $\Si^0\Si^0$, $\La \La$).
The interesting point is that scenario (1) and scenario (2) are 
indistinguishable.
Both look like the ``CFL'' phase of the figure:
$U(1)_S$ and chirality are broken, and there is an
unbroken vector $SU(2)$. This is the ``continuity of quark and hadron matter''
described by Sch\"afer and Wilczek \cite{SchaeferWilczek}.
We conclude that for low enough strange quark mass, $\ms<\ms^{\rm cont}$, there
may be a region where sufficiently dense baryonic matter has the same
symmetries as quark
matter, and there need not be any 
phase transition between them. In Section 6 we use this observation to
construct a mapping between 
the gaps we have calculated at the Fermi surfaces in
the quark matter phase  and gaps at the baryonic
Fermi surfaces at lower densities.

All of the
qualitative features shown in Figure 1
follow from the above discussion of the small and large $m_s$
regimes, with one exception. In Figure 1, a transition between
two flavor nuclear matter and three flavor quark matter (in the 2SC+s
phase) occurs only for a range of values of $m_s$ above $m_s^{\rm cont}$.
However, it may be that the strange baryon
phase ends {\it below} $m_s^{\rm cont}$. One would then have 
a transition
between two flavor nuclear matter and three flavor quark
matter (in either the 2SC+s phase or the CFL phase)
for a range of $m_s$ values which extends both above and below
$m_s^{\rm cont}$.  
Determining
the extent of the strange baryon phase requires a detailed
analysis of strange baryonic matter, including the possibility
of kaon condensation. This is not our goal here.

This concludes our overview of the 
qualitative features of the phase diagram for low temperature
strongly interacting matter.  In subsequent sections
we analyze the 2SC and CFL phases and 
the unlocking transition more quantititatively,
and use the properties of the CFL phase to make
predictions for properties of baryonic
matter in which $U(1)_S$ is broken while 
$SU(2)_V$ is unbroken.

\section{Model Independent Features of the Unlocking Phase Transition}
\label{sec:general}

In this section, we give a model independent argument that the 
unlocking phase transition between the CFL and 2SC phases
in Figure 1 must be first order.

For any $m_s\neq 0$, transformations in $SU(3)_A$ which
involve the strange quark are explicitly not symmetries.
The CFL and 2SC phases are distinguished by whether
the chiral $SU(2)_A$ rotations involving
only the $u$ and $d$ quarks are or are not spontaneously
broken.  As we will make clear in subsequent
sections, the unlocking transition
is associated with the vanishing of those diquark
condensates which pair a strange quark with either an up or
a down quark.  We denote the
resulting gaps $\Delta_{us}$ for simplicity.
In the absence of any $\Delta_{us}$ gap,
the only
Cooper pairs are those involving pairs of
light quarks, or pairs of strange quarks.  
The light quark
condensate is unaffected by the strange quarks, and
behaves as in a theory with only two flavors of
quarks. Chiral symmetry is unbroken.  
We will show explicitly below that when $\Delta_{us}\neq 0$,
the interaction between the light quark condensates and
the mixed (light and strange) condensate results in the 
breaking of two-flavor chiral symmetry $SU(2)_A$ via the locking of
$SU(2)_L$  and $SU(2)_R$ 
flavor symmetries to an $SU(2)$ subgroup of color.  This
color-flavor locking mechanism leaves a global
$SU(2)_{{\rm color}+V}$ group unbroken.

\begin{figure}[t]
\epsfxsize=3.5in
\begin{center}
\hspace*{0in}
\epsffile{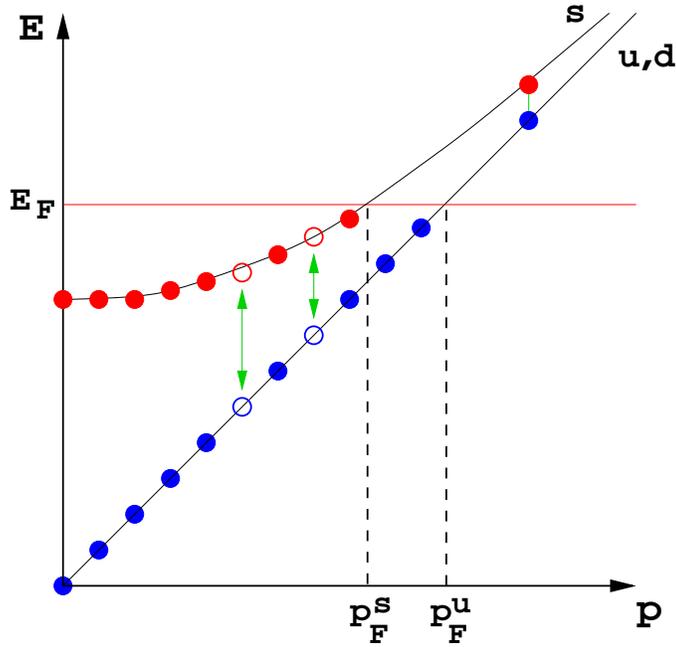}
\end{center}
\caption{
How the strange quark mass disrupts a $u$-$s$ condensate.  The strange
quark (upper curve) and light quark (straight line) dispersion
relations are shown, with their Fermi seas filled up to the Fermi
energy $E_F$.  The horizontal axis is the magnitude of the
spatial momentum; pairing occurs between particles (or holes)
with the same $p$ and opposite $\vec p$.
For $p<p_F^s$, hole-hole pairing ($\bar s$-$\bar u$) is
possible (two examples are shown).  For $p>p_F^u$, particle-particle
pairing ($s$-$u$) is possible (one example is shown).  Between the
Fermi momenta for the $s$ and $u$ quarks, no such pairing is possible.}
\label{fig:fermi}
\end{figure}

The unlocking transition is a transition 
between a phase with $\Delta_{us}\neq 0$ at  $M_s<M_s^{\rm unlock}$
and a phase with
$\Delta_{us}=0$.  ($M_s(\mu)$
is the constituent strange quark mass at chemical potential $\mu$.)
The BCS mechanism guarantees superconductivity in the 
presence of an arbitrarily weak attractive interaction,
and there is certainly an attraction between $u$ and $s$
quarks (with color ${\bf \bar 3}$) for any $M_s$.
How, then, can $\Delta_{us}$ vanish above $M_s^{\rm unlock}$?
The BCS result relies on a singularity
which arises for pairs of fermions with zero total momentum
{\it at} the Fermi surface.
We see from Figure 2 that no pairing is possible for quarks
with momenta between the $u$ and $s$ Fermi momenta, and that
at most
one of the quarks in a $u$-$s$ Cooper pair
can be at its respective Fermi surface.
The BCS singularity therefore does not arise if $M_s\neq 0$, 
and a $u$-$s$ condensate is not mandatory.  
A $u$-$s$ condensate involves pairing of quarks with momenta
within about $\Delta_{us}$ of the Fermi surface, 
and we therefore expect that
$\Delta_{us}$ can only be nonzero if the mismatch
between the up and strange Fermi momenta is less than
or of order $\Delta_{us}$:
\begin{equation}
\sqrt{\mu^2-M_u(\mu)^2} - \sqrt{ \mu^2-M_s(\mu)^2} 
 \approx \frac{M_s(\mu)^2-M_u(\mu)^2}{2\mu} 
\lesssim \Delta_{us}\ .
\label{criterion}
\end{equation}
Here $M_s(\mu)$ and $M_u(\mu)$ are the constituent quark masses
in the CFL phase. We neglect $M_u(\mu)$ in the following.
Equation (\ref{criterion}) 
implies that arbitrarily small values of $\Delta_{us}$ are
impossible.
As $m_s$ is increased from zero, $\Delta_{us}$ decreases
until it is comparable to $M_s(\mu)^2/2\mu$. At this point, 
smaller nonzero values of $\Delta_{us}$ are not possible, 
and $\Delta_{us}$  must therefore
vanish discontinuously.  This simple
physical argument leads us to conclude that the
unlocking phase transition at $M_s=M_s^{\rm unlock}$
must be first order.

Below, we confirm by
explicit calculation in a model that the
unlocking phase transition is first order.
We find that $\Delta_{us}$ is of order
$50-100~\MeV$ on the CFL side 
of the transition if the coupling is calibrated to give a
reasonable magnitude for the
chiral condensate in vacuum.

\section{Superconducting Condensates in a Model}
\label{sec:con}

We study the physics of the quark matter phases in Figure 1
in a toy model
in which we replace the full
interactions between quarks by a four-fermion interaction
with the quantum numbers of single-gluon exchange,
\begin{equation}
{\cal L}_{\rm interaction} = G\int d^4 x \left( \bar q \lambda^A 
\gamma^\mu q\right)\left( \bar q \lambda^A 
\gamma_\mu q\right)\ ,
\label{con:interaction}
\end{equation}
and work in a mean-field approximation.
All four-fermion interactions involving fermions at the Fermi surface
are equally relevant, in the renormalization group sense, so
why use only the one with the quantum numbers of one-gluon exchange?
Renormalization group analyses\cite{SchaeferWilczekRG,EHS} show that other
interactions are important, but confirm that in QCD with two and three
massless quarks the most attractive
channels for condensation are those generated by (\ref{con:interaction}).
There is one important caveat.
The single-gluon-exchange interaction is symmetric under
$U(1)_A$, and so it sees no distinction between
condensates of the form $\<q C q \>$ and $\<q C \ga_5 q \>$.
However, once instantons are included the Lorentz scalar
$\<q C \ga_5 q \>$ is favored,\cite{ARW2,RappETC} 
so we neglect the other form.

One of the 
central qualitative lessons of past work is that superconducting
gaps of order 100 MeV are obtained in
the quark matter phase independent of the details
of the interaction which is used, as long as the strength
of the interaction is chosen in a way which is roughly
consistent with what we know about the vacuum chiral condensate.  
Superconducting
gaps of this magnitude can be obtained using a number of different
treatments based upon single-gluon exchange.\cite{BailinLove, IwaIwa, ARW3}
Refs.~\cite{ARW2, RappETC, BergesRajagopal, CarterDiakonov} find gaps
of this magnitude 
by making approximations of varying sophistication in which
the interaction between
quarks is modeled by 
that induced by instantons.  Further confirmation of this
lesson comes from recent work using the NJL model.\cite{Klevansky}

As in any model which uses a four-fermion interaction, 
in addition to choosing the quantum numbers of the interaction
we must introduce an ad hoc form factor.
The explicit lesson of Refs.~\cite{ARW2,ARW3,Klevansky},
implicit in other work, is that results are rather insensitive to
the choice of form factor, again
as long as $G$ is suitably normalized.
We therefore 
use a simple sharp cutoff in momentum integrals at 
a spatial momentum $\Lambda=800$~MeV.
Our results should therefore not be extended above $\mu\sim\Lambda$.  
The form factor (which we
describe by the single parameter $\Lambda$) is a crude
representation of physics not described 
by the model, and $\Lambda$ should not be taken to infinity.  
For $\Lambda=0.8$ GeV, setting
$G=7.5$ GeV$^{-2}$ in (\ref{con:interaction}) 
results in a vacuum constituent mass for the
light quarks of 400 MeV, and we use these values 
of $\La$ and $G$ throughout.
We have however checked that if we instead take $\Lambda=1.$ GeV,
and change $G$ accordingly, the superconducting gaps and
the location of the unlocking phase transition are not
significantly affected.

An analysis of the phase diagram requires
condensates
\begin{equation}
\label{con:conds}
\<q^\al_i C\ga_5 q^\be_j\>\ ,\ \ \ 
\<q^\al_i C\ga_5\ga_4 q^\be_j\>\ ,\ \ \
\< \bar q^{\,i}_\al q^\be_j \>\ ,
\end{equation}
leading to gap parameters
\begin{equation}
\label{con:gaps}
\De^{\al\be}_{ij}\ ,\ \ \ \ \ \ \ \ \ \ 
\ka^{\al\be}_{ij}\ ,\ \ \ \ \ \ \ \ \ \ 
\phi^{i\be}_{\al j}
\end{equation}
with corresponding quantum numbers.  Each
gap matrix is a 
symmetric $9\times 9$ matrix describing the color (Greek indices)
and flavor (Roman indices) structure.  

In \ref{sec:appendix} we derive the gap equation for these condensates
in full generality, assuming nothing about their color-flavor
structure.  However, it is very difficult to solve the general gap
equation, so at this point we make several simplifying assumptions to
obtain easily soluble gap equations.  These are:
\begin{enumerate}
\item We fix $\phi^{i\be}_{\al j} = \Ms \de^i_3 \de^3_j \de_\alpha^\beta$.
For more details see below.
\item We use the simplest form for $\De^{\al\be}_{ij}$ which
allows an interpolation between the color-flavor locking favored
by single-gluon exchange at $m_s=0$ and the ``2SC'' phase
favored at $m_s\to\infty$ (see \eqn{sol:ansatz}).
This ansatz, which requires
five independent superconducting gap parameters, leads to 
consistent gap equations in the presence of the one-gluon
exchange interaction. We describe this ansatz in detail
in Section 4. 
\item We neglect $\ka^{\al\be}_{ij}$. This condensate
pairs left-handed and right-handed quarks, and so breaks
chiral symmetry. It can be shown to vanish in the absence
of quark masses.\cite{PisarskiRischke}
In \ref{sec:appendix}
we show that it {\it must} be nonzero in the presence 
of a nonzero $\langle q C \gamma_5 q \rangle$ condensate and 
nonzero quark masses. In the 2SC phase, the 
$\langle q C \gamma_5 q \rangle$ condensate involves only the massless
quarks and $\langle q C \gamma_5 \gamma_4 q \rangle$ 
therefore vanishes and chiral symmetry remains unbroken.
In the CFL phase, however, the 
$\langle q C \gamma_5 q \rangle$ condensate involves the 
massive strange quarks and itself breaks chiral
symmetry by color-flavor locking.
No symmetry argument precludes $\kappa\neq 0$, and in
\ref{sec:appendix} we have confirmed by direct calculation at one 
value of $\mu$
that the gaps $\ka^{\al\be}_{ij}$ are nonzero in the CFL phase.
However, we find that these gaps are much
smaller than the corresponding gaps generated by the $\langle q C
\gamma_5 q \rangle$ condensate. Including $\ka$ in the gap equations
for $\De$ 
makes finding solutions prohibitively slow.
Therefore, once we have convinced ourselves that these condensates are
present but small, we neglect them.
\end{enumerate}

We now give a more detailed discussion of
our assumptions about the quark mass matrix $\phi$.
Since all our calculations are in the quark matter regime ($\mu>\mu_V$
in Figure~\ref{fig:phasediagram}), we will assume that $\mu$ is high
enough that we can neglect 
chiral condensates for the $u$ and $d$ quarks,\footnote{
Because chiral symmetry is broken in the CFL phase, 
there is no reason for the ordinary $\langle \bar u u \rangle$ and
$\langle \bar d d\rangle$ chiral condensate to vanish. Indeed,
Ref.~\cite{ARW3} demonstrates that the $\langle \bar u u \rangle$ and
$\langle \bar d d\rangle$ condensates must be nonvanishing.  However,
an explicit calculation \cite{Schaefer} shows them to be small, 
and we neglect them. \Eqn{criterion} shows that nonzero $\<\bar u u\>$
and $\<\bar d d\>$ condensates allow color-flavor locking to persist
to higher strange quark masses.}
but
because the current quark mass $m_s$ is nonzero there may
be a nonzero $\langle \bar s s \rangle$ chiral condensate.
In this case the quark mass matrix
is $\phi={\rm diag}(0,0,\Ms)$, where $\Ms$ is the $\mu$-dependent
constituent strange quark mass, satisfying a gap equation of its
own. At sufficiently high densities, it is given by the current mass
$\ms$, of order 100 MeV.  At lower densities, it receives an
additional contribution from $\langle \bar s s \rangle$.
In the interests of simplicity, however, we will not solve the $\Ms$
gap equation simultaneously with the superconductivity gap equations.
Instead, we treat $\Ms$ as a parameter in the 
superconductivity gap equations, and do not 
determine what value of $\Ms$ corresponds to given values
of $\ms$ and $\mu$. For simplicity, we choose the
same value of the parameter $\Ms$ for strange quarks
of all three colors.  This would not be the case in a 
treatment in which the $\Ms$ gap equations are solved
simultaneously with the superconductivity gap 
equations.\cite{CarterDiakonov}

With our simplifying assumptions, we are left needing only a gap equation for
the quark-quark condensate $\De^{\al\be}_{ij}$.
The simplest ansatz that
interpolates between the two flavor case ($\ms=\infty$)
and the three flavor case ($\ms=0$) is
\addtocounter{equation}{3}
\beql{sol:ansatz}
%\begin{equation}
\ba{l}
\De^{\al\be}_{ij} =
\left(
\ba{ccccccccc}
b+e & b & c \\
b & b+e & c \\
c & c & d \\
  &   &   & & e \\
  &   &   & e & \\
  &   &   & & & & f\\
  &   &   & & & f &\\
  &   &   & & & & & & f\\
  &   &   & & & & & f &\\
\ea
\right) \\
\hbox{basis vectors:} \\
\ba{rcl@{\,\,}l@{\,\,}l@{\,\,\,\,}
      l@{\,\,}l@{\,\,\,\,} l@{\,\,}l@{\,\,\,\,}l@{\,\,}l}
(\al,i) &=& (1,1),&(2,2),&(3,3),&(1,2),&(2,1),&(1,3),&(3,1),&(2,3),&(3,2) \\
        &=& (r,u),&(g,d),&(b,s),&(r,d),&(g,u),&(r,s),&(b,u),&(g,s),&(b,d)
\ea
\ea
%\label{sol:ansatz}
%\end{equation}
\eeql
where the color indices are $\al,\be$ and the flavor indices are $i,j$.
The strange quark is $i=3$. The rows are labelled by $(\al,i)$ and the
columns by $(\be,j)$.

\begin{table}[hbt]
\def\st{\rule[-1.5ex]{0em}{4ex}} 
\begin{center}
\begin{tabular}{llll}
\hline
\st description & condensate & symmetry  \\
\hline
$\ba{l}\hbox{2SC: 2-flavor} \\ \hbox{\phantom{2SC: }superconductivity}\ea$
 & $\ba{l} c=d=f=0,\\ b=-e \ea$
     & $SU(2)_L\times SU(2)_R$  \\[2ex]
$\ba{l} \hbox{CFL: color-flavor locking}\ea$ &  &   
$SU(2)_{{\rm color}+L+R}$   \\[0.5ex]
$\ba{l}\hbox{CFL: color-flavor locking} \\ \hbox{\phantom{CFL: }with $\ms=0$}\ea$
& $\ba{l} c=b,f=e,\\ d=b+e \ea $ 
     & $SU(3)_{{\rm color}+L+R}$ \\
\hline
\end{tabular}]
\end{center}
\caption{
Symmetries of the condensate ansatz \eqn{sol:ansatz} in various
regimes.
}
\label{tab:syms}
\end{table}
The properties of the ansatz are summarized in Table \ref{tab:syms}.
In its general form, this condensate locks color and flavor.
This is because of the condensates $c$ and $f$, referred to 
collectively as $\Delta_{us}$ above,
that
combine a strange quark with a light one.
It is straightforward
to confirm by direct calculation that if either $c$ or $f$ or $b+e$ is
nonzero, then the matrix $\De^{\al\be}_{ij}$ of (\ref{sol:ansatz}) is
not invariant under separate flavor or color rotations but is 
left invariant by simultaneous rotations of $SU(2)_V$ and the $SU(2)$
subgroup of color corresponding to 
the colors 1 and 2.  Thus, color-flavor locking occurs whenever  one
or more of $c$, $f$, or $b+e$ is nonzero.
We discuss $b+e$ below.

Although the standard electromagnetic symmetry is broken
in the CFL phase, as are all the color gauge symmetries,
there is a combination of electromagnetic and color
symmetry that is preserved.\cite{ARW3} Consider
the gauged $U(1)$ under which the charge $Q'$
of each quark is the sum of its electromagnetic 
charge $(2/3,-1/3,-1/3)$ (depending on the flavor
of the quark) and its color hypercharge $(-2/3,1/3,1/3)$ (depending
on the color of the quark).  It is easy to confirm that the 
sum of the $Q'$ charges of each pair of quarks corresponding
to a nonzero entry in (\ref{sol:ansatz}) is zero. This modified
electromagnetism is therefore not broken by the condensate.

At $M_s=0$, one has color-flavor locking: $c$, $f$ and $b+e$ 
are all
nonzero. In addition, 
$c=b$, $f=e$, and $d=b+e$, and
the matrix is invariant under simultaneous
rotations of $SU(3)_V$ and $SU(3)_{\rm color}$.
For $M_s$ nonzero but sufficiently small, $c$, $f$ and $b+e$
remain nonzero but are no longer equal to $b$, $e$ and $d$.
Color and flavor are locked, and the matrix is invariant
under simultaneous
rotations of $SU(2)_V$ and $SU(2)_{\rm color}$.
As described in Section 2,
once $\Ms$ becomes large enough, $c$ and $f$ both 
vanish (and so does $b+e$) and the symmetry
group enlarges, unlocking color and flavor, and restoring chiral symmetry
(see Table \ref{tab:syms}). 

If $\Delta$ were nonzero only in color
${\bf \bar 3}$ channels, we would have $b=-e$, $d=0$ and $c=-f$. 
Single-gluon-exchange is repulsive in 
the color ${\bf 6}$ channels.
However, in the CFL phase where $c$ and $f$ are
nonzero, the only consistent solutions to the gap 
equations have $b\neq -e$, $c\neq -f$, and $d\neq 0$.
This means that the single-gluon-exchange interaction {\it requires}
a small color ${\bf 6}$ admixture
along with the favored color ${\bf \bar 3}$ condensate, even
though it prohibits color ${\bf 6}$ condensates alone.
This feature
occurs both for $M_s=0$\cite{ARW3} and for $0< M_s < M_s^{\rm unlock}$.
Note that not all color ${\bf 6}$
terms arise in (\ref{sol:ansatz}). For example, only three of the nine 
diagonal elements of (\ref{sol:ansatz}) are nonzero.
These are the only three diagonal elements 
which can be nonzero without
breaking the $SU(2)_{{\rm color}+V}$
symmetry, and upsetting color-flavor locking. 
Furthermore, if the two upper-most diagonal elements were not
equal to $b+e$, the $SU(2)_{{\rm color}+V}$ symmetry
would also be broken.
We have therefore learned that
the color ${\bf 6}$ terms which are induced in the presence of a 
color ${\bf \bar 3}$ condensate
are those which do not change the global symmetry of the 
color ${\bf \bar 3}$ condensate. Those color ${\bf 6}$ terms
which are ``allowed'' in this sense are generated by the interaction.

The corresponding analysis of the 2SC phase,  
where only two flavors participate in the 
condensate is similar in logic but
leads to different conclusions.  
In the 2SC phase, where $c$ and $f$ are zero, we do in fact find $b=-e$
and $d=0$, namely a pure color ${\bf \bar 3}$ condensate.
In this case,
if $b+e$ were nonzero, one {\em would} have 
color-flavor locking and broken chiral symmetry.
In this case, then, the pure color ${\bf \bar 3}$
condensate with $b+e=0$ is ``protected'' by the fact
that any admixture of color ${\bf 6}$ condensate like $b+e$
would break a global symmetry which the color ${\bf \bar 3}$ condensate
does not break.
In the absence of 
$c$ and $f$, a nonzero $d$ condensate ($\langle ss \rangle$)
is still allowed, but there is no way for the 
color ${\bf \bar 3}$ condensate to induce it. Therefore the simple
argument that one-gluon exchange is repulsive in this channel
holds, and in our model 
we find $d=0$ in the 2SC phase even where the strange quark density
is nonzero.
As discussed above, a more complete analysis would include
a small $\langle ss \rangle$ condensate in some channel, resulting
in a distinction between the 2SC and 2SC+s phases which our analysis
does not see.

\section{The Gap Equation \ldots }
\label{sec:sol}

Our task now is to obtain the gap equation for $\De$,
and verify the
behavior described in Sections 1 and 2. In particular, we want to
show that the unlocking phase transition is first-order, and
calculate $\Ms^{\rm unlock}(\mu)$ and compare 
to the criterion (\ref{criterion}) which we derived
on model-independent grounds.

In \ref{sec:appendix} we have derived the 
mean field gap equation and given its general
solution. It takes the form (see \Eqn{app:gap1})
\beql{sol:gap1}
\De = {G\over (2\pi)^4} \int \! d^4q \,
\la^T_A \ga_\mu \,P_1(\mu,\Ms,\De,q) \,\la_A \ga^\mu,
%\label{sol:gap1}
\eeql
where the function $P_1$ is given by \eqn{app:X} and \eqn{app:Xinv},
and we are assuming, as described in Section~\ref{sec:con} that 
$\phi^{i\be}_{\al j} = \Ms \de^i_3 \de^3_j \de_\alpha^\beta$
and that $\ka=0$. Note that strictly speaking the gap equations do not
close for $\ka=0$. The $\ka$ gap equation is just like \eqn{sol:gap1}
but with $-P_4$ instead of $P_1$ on the right hand side, and for $\Ms>0$,
$P_4$ is nonzero even when $\ka=0$. This means that it is
inconsistent to set $\ka=0$.
However, as discussed in the Appendix,
$\ka$ turns out to be small and we use (\ref{sol:gap1}) with 
$\ka=0$ in $P_1$.

In principle we could just evaluate $P_1(\mu,\Ms,\De,q)$
by substituting  \eqn{sol:ansatz} into \eqn{app:Xinv}, but that would
lead to a very complicated expression.
To turn \eqn{sol:gap1} into a tractable set of gap equations, we transform
to a slightly different basis from \eqn{sol:ansatz}.
In the new basis, the first 2 vectors are 
$(\al,i)= 1/\sqrt{2}\Bigl( (1,1)\pm(2,2) \Bigr)$.
Now the top left $3\times 3$ block of \eqn{sol:ansatz} looks like
\beq
\left(
\ba{ccc}
e \\
& a_{11} & a_{12} \\
& a_{12} & a_{22} 
\ea
\right)
=
\left(
\ba{ccc}
e \\
& 2b+e & \sqrt{2}c \\
& \sqrt{2}c & d 
\ea
\right)
\eeq
$\De$ is now block-diagonal, consisting of $1\times 1$ and $2\times 2$
blocks. We find that
$P_1(\mu,\Ms,\De,q)$ takes a corresponding 
block-diagonal form,
\beq
\ba{l}
P_1(\mu,\Ms,\De,q) = \\[2ex]
\left(
\ba{c@{\!\!\!\!\!}c@{\,\,}c@{\!\!\!}c@{\!}c@{\!\!\!}c@{\!}c@{\!\!\!}c@{\!}c}
E(e,q)  \\
 & A_{11}(a,q) & A_{12}(a,q) \\
 & A_{12}(a,q) & A_{22}(a,q) \\
  &   &   & & E(e,q) \\
  &   &   & E(e,q) & \\
  &   &   & & & & F(f,q)\\
  &   &   & & & F(f,q) &\\
  &   &   & & & & & & F(f,q)\\
  &   &   & & & & & F(f,q) &\\
\ea
\right) \\
\ea
\eeq
where 
\beql{sol:RHS}
\ba{r@{\,\,}c@{\,\,}l@{\,\,}c@{\,\,}l}
F(f,q) &=& P_1(\mu,\Ms,f,q) &=& \dsp
{f w \over w^2 - (4\mu^2-\Ms^2)\,\qvsq - (\mu- iq_0)^2 \Ms^2 } \\[2ex]
&& \phantom{ P_1(\mu,\Ms,q)}(w &=& f^2 + \mu^2 + \qvsq + q_0^2) \\[0.5ex]
E(e,q) &=& P_1(\mu,\,0\,,e,q) &=& 
  \hbox{as above, with $f\to e$ and $\Ms=0$} \\[0.5ex]
A(a,q) &=& P_1(\mu,\Ms,a,q)
\ea
\eeql
Note that the function $P_1$ simplifies considerably when its argument
is a symmetric off-diagonal $2\times 2$ matrix, as in the $e$ and $f$
blocks. The resulting expressions for $E$ and $F$ are given on the
right-hand side of \eqn{sol:RHS}.
The only difficult thing in \eqn{sol:RHS} is evaluating $P_1$
for the $2\times 2$ matrix $a$, in which case Eqns.~\eqn{app:X} and
\eqn{app:Xinv} must be used.

The only nonzero matrix elements in $P_1$,
on the right hand side of the gap equation, are those
which are nonzero in our ansatz for $\Delta$, on the left
hand side of the gap equation.  This confirms that the
only color ${\bf 6}$ terms which are induced by the 
color ${\bf \bar 3}$ condensate are those which we 
have included in our ansatz, in agreement with the
symmetry arguments given above.  Although more complicated
ans\"atze may be worth considering in future work, 
this ansatz is the minimal one which suffices.

Taking into account the $\la$ and $\ga$ matrices from the gluon vertex,
we end up with the 5 gap equations
\beql{sol:gap2}
\ba{rcl}
e &=&  {4G/ (2\pi)^4} \int\!d^4q\,\, 
     \Bigl(A_{11}(a,q) - {5\over 3} E(e,q) \Bigr) \\[1ex]
f &=&  {4G/ (2\pi)^4}  \int\!d^4q\,\,
    \Bigl(\sqrt{2}A_{12}(a,q) - {2\over 3} F(f,q) \Bigr)\\[1ex]
a_{11} &=& {4G/ (2\pi)^4} \int\!d^4q\,\,
    \Bigl({1\over 3} A_{11}(a,q) + 3 E(e,q) \Bigr)\\[1ex]
a_{12} &=& {4G/ (2\pi)^4} \int\!d^4q\,\,
   \Bigl(-{2\over 3} A_{12}(a,q) + 2\sqrt{2} F(f,q) \Bigr)\\[1ex]
a_{22} &=& {4G/ (2\pi)^4} \int\!d^4q\,\,
    {4\over 3} A_{22}(a,q)  
\ea
\eeql

To obtain the gap parameters as a function of $\mu$ and $\Ms$, we solve the
5 simultaneous equations \eqn{sol:gap2} numerically, using 
{\em Mathematica}. The right-hand sides are evaluated by numerical
integration of the integrands given in \eqn{sol:RHS}.
For $A(a,q)$, we must evaluate $P_1(\mu,\Ms,a,q)$, which corresponds to
\eqn{app:Xinv} with $\De$ being the $2\times 2$ matrix $a$,
$\ka=0$, and $\phi=\hbox{diag} (0,\Ms)$. The gap parameters $b,c,d,e,f$
determine the poles of the propagator $P_1$ and hence
determine the quasiparticle dispersion relations and the gaps
at the Fermi surface.

\section{\ldots and Its Solutions}
\label{sec:res}

\begin{figure}[t]
\epsfxsize=4in
\begin{center}
\hspace*{0in}
\epsffile{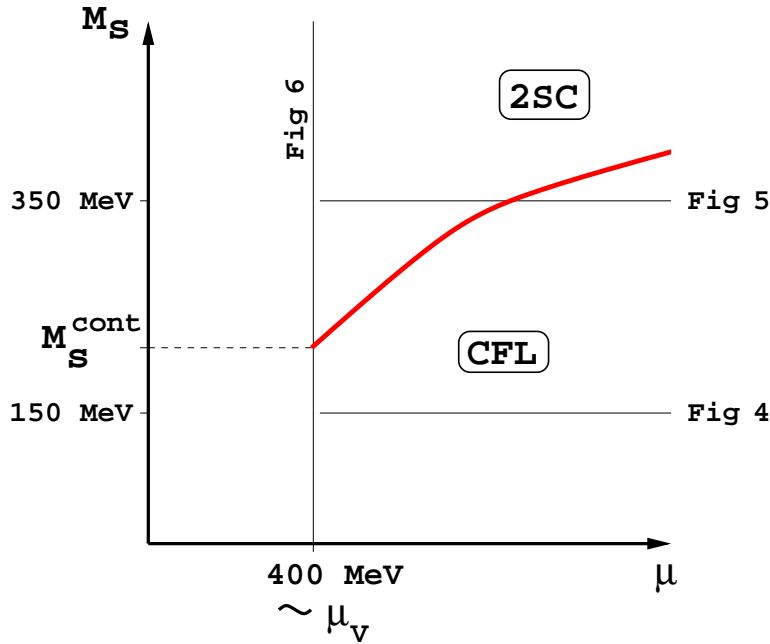}
\end{center}
%\vspace{-10ex}
\caption{
The part of the phase diagram in which we have 
done calculations using our model. 
The thin solid lines show the slices along which
we have solved the gap equations. Note that a horizontal
line in Figure 1 at some $\ms$
corresponds to a curve in this figure which
approaches the horizontal line $M_s=m_s$ from above as
$\mu\rightarrow\infty$. 
}
\label{fig:pd2}
\end{figure}
In this section, we present solutions to the gap equations
\eqn{sol:gap2}. We demonstrate explicitly that the 
unlocking phase transition is first order in our model,
as we have argued on general grounds in Section 2.
Furthermore, we confirm that the simple criterion
(\ref{criterion}) is a good guide.
We determine the 
value of $\Ms^{\rm cont}$, the highest strange quark constituent 
mass at which, in the model, no 2SC phase intrudes.
We use our results to draw tentative conclusions for
the behavior of strongly interacting matter as a 
function of density in nature.

We plot the solutions $b,c,d,e,f$ to the gap equations along two
different
lines of constant $M_s$ and one line of constant $\mu$, all shown
in Figure~\ref{fig:pd2}. Recall that $b=-e$ 
and $c=d=f=0$ in the 2SC phase whereas all the gaps are nonzero
in the CFL phase.  For $\Ms=0$, $b=c$ and $e=f$.  
If the condensate were purely color ${\bf \bar 3}$ 
in the CFL phase, one would have $b=-e$, $c=-f$ and $d=0$.

\begin{figure}[t]
\epsfxsize=4.5in
\begin{center}
\hspace*{0in}
\epsffile{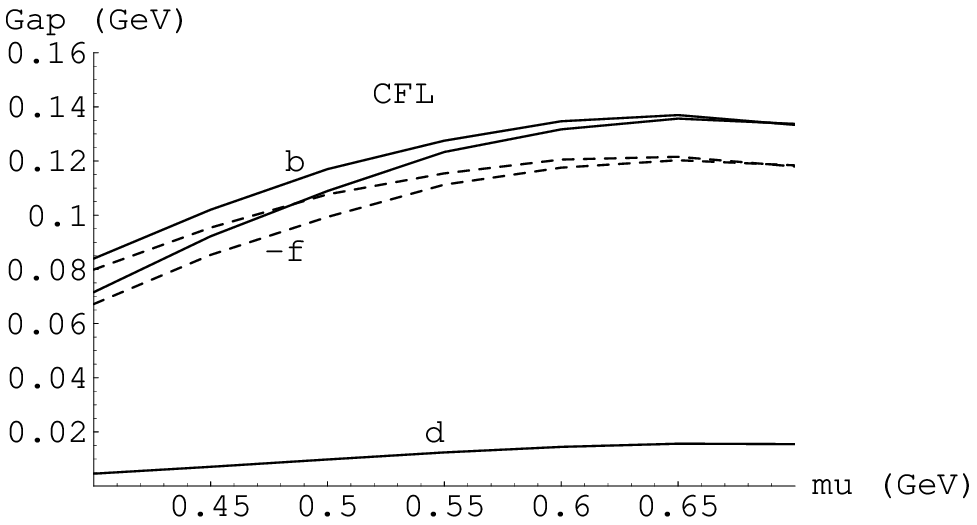}
\end{center}
\vspace{-15ex}
\caption{
Solutions to the gap equations
as a function of $\mu$, at $\Ms=150~\MeV$. At this value of $\Ms$, the
CFL phase persists down to $\mu=400~\MeV$, and below.
The two solid lines are $b$ and $-e$, the dashed lines are 
$c$ and $-f$. At large $\mu$, the gaps approach the $M_s=0$ pattern.
}
\label{fig:gap-ms0.15}
\end{figure}
\begin{figure}[hbt]
\epsfxsize=4.75in
\begin{center}
\hspace*{0in}
\epsffile{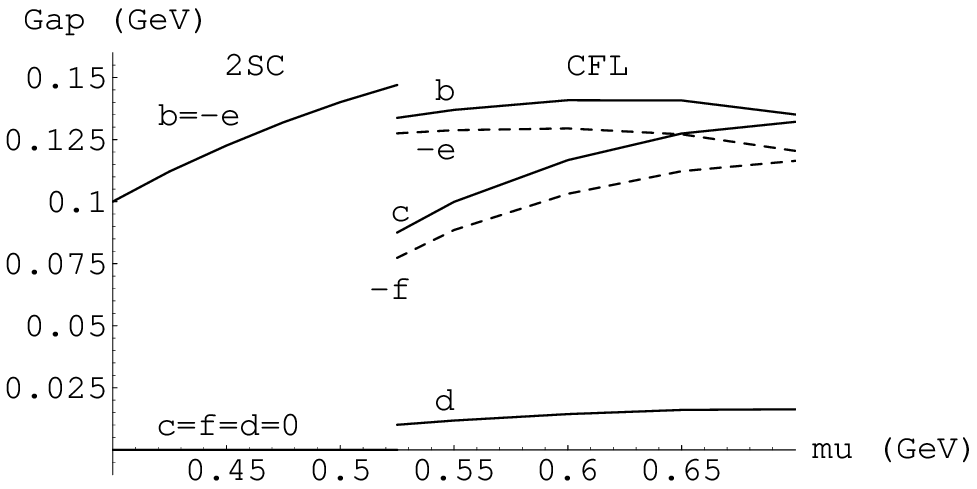}
\end{center}
\vspace{-15ex}
\caption{
Solutions to the gap equations
as a function of $\mu$, at $\Ms=350~\MeV$.
At this large value of $\Ms$, the CFL phase exists only for 
$\mu> 530~\MeV$. That is, $M_s^{\rm unlock}=350~\MeV$ for $\mu=530~\MeV$.
}
\label{fig:gap-ms0.35}
\end{figure}
Figures 4 and 5 show the solutions to the gap equations 
as a function of $\mu$,
for $M_s=150~\MeV$ and $M_s=350~\MeV$.  
We see that for $M_s=150~\MeV$,
the color-flavor-locked phase continues down to 
$\mu=400~\MeV$, below which we expect a baryonic description
to become appropriate.
In contrast, in Figure 5 we see that for $M_s=350~\MeV$, 
the CFL phase only exists for $\mu>530~\MeV$. At lower
densities, below a first order unlocking phase transition,
the 2SC phase is favored.\footnote{In the calculations
for our plots we assume $\Ms$ is the same on either side of 
the phase transition. In reality $\Ms$ will be somewhat smaller in the 2SC
phase because there is no chiral symmetry breaking there, but since
the 2SC condensates do not involve the $s$ quark this is of no
consequence.}
This transition occurs at high 
enough densities that a quark matter description is justified,
and our model can therefore be used to describe it.
At lower densities
still, there is another first order phase transition
to a baryonic 
phase.\cite{ARW2,RappETC,BergesRajagopal,BJW2,PisarskiRischke1OPT,CarterDiakonov}

\begin{figure}[htb]
\epsfxsize=4.75in
\begin{center}
\hspace*{0in}
\epsffile{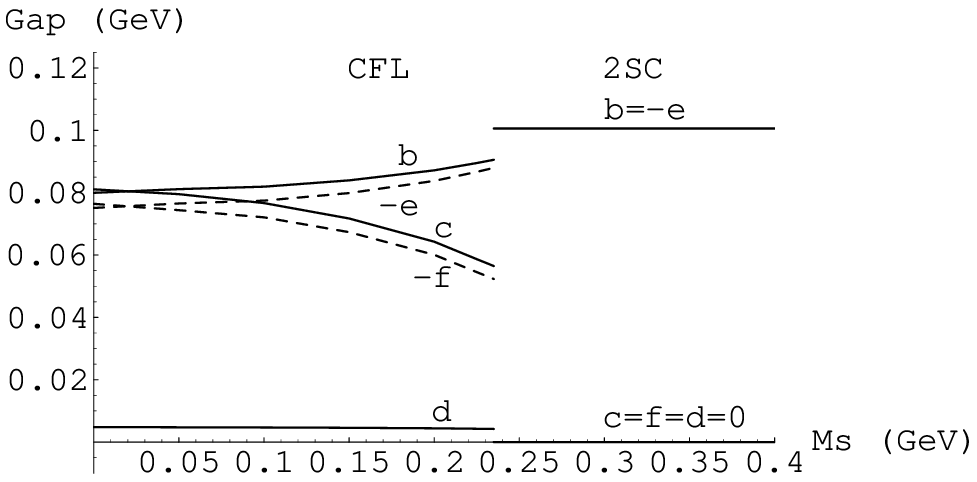}
\end{center}
\vspace{-15ex}
\caption{
Solutions to the gap equations
as a function of $\Ms$, at $\mu=400~\MeV$, along the vertical
line in Figure 3.
When the strange quark becomes massive enough, $s$-$u$ condensates
become too expensive, and there is a first-order phase transition from
CFL to 2SC.
}
\label{fig:gap-mu0.4}
\end{figure}
In order to estimate $M_s^{\rm cont}$, in Figure 6 we plot
the solutions to the gap equations as a function of $M_s$ with 
$\mu$ fixed at $400~\MeV$.
This is the lowest $\mu$ at which we expect a quark matter
description to be valid.
We find that the CFL phase
exists for $M_s$ below a first order transition at 
$M_s^{\rm unlock}=235~\MeV$.  Our model therefore suggests that 
$M_s^{\rm cont}$ is of order $250~\MeV$.

The question, then, is whether Figure 4 or Figure 5 is
closer to the behavior of strongly interacting matter
as a function of density in nature.   If Figure 5 shows
the correct qualitative behavior, we would expect two first order phase
transitions as the density is increased above nuclear
density. After the first transition, one would have
quark matter in the 2SC phase. In this quark matter
phase, the constituent strange quark mass must 
be greater than $M_s^{\rm cont}\sim 250~\MeV$.  Then,
above the next  transition, a color-flavor-locked
quark matter phase is obtained.  
If Figure 4 is the better
gu1ide to nature, as we surmise, then there is no
window of $\mu$ in which the 2SC phase intrudes, and
nature may choose a continuous transition between baryonic
and quark matter with the symmetries of the CFL phase.

\begin{figure}[htb]
\epsfxsize=4in
\begin{center}
\hspace*{0in}
\epsffile{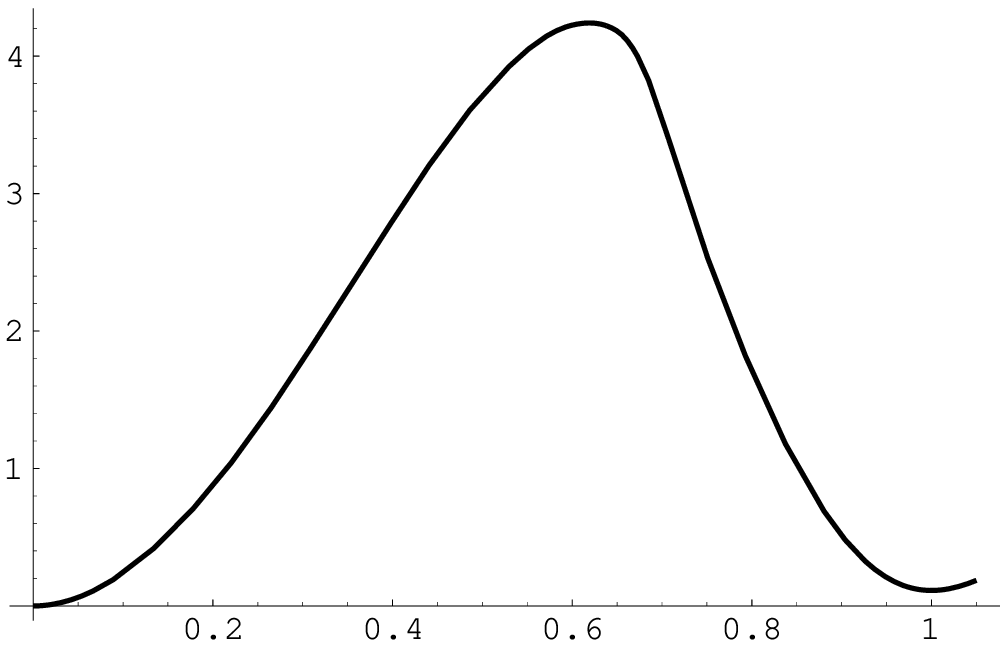}
\end{center}
\vspace{-0ex}
\caption{
The effective potential $\Omega$ in units 
of $10^{-5}~\GeV^4$ in the vicinity 
of the unlocking phase 
transition, for $\mu=400~\MeV$, $\Ms=235~\MeV$. The plot corresponds
to a slice of the potential in the 5 dimensional parameter 
space along a line which linearly interpolates
between the minima in the 2SC phase ($x=0$) and 
the CFL phase ($x=1$). 
The phase transition is clearly first order.
}
\label{fig:firstorder}
\end{figure}
We have analyzed the unlocking phase transition quantitatively
in our model, and can now confirm the predictions
of the model-independent but qualitative arguments of Section 2.
We first demonstrate that the transition is first order,
and justify the location of the discontinuities
shown in Figures 5 and 6.
To find the first order transition in our model, we need 
the effective potential $\Omega$.
We use the fact that the gap equations  
are equivalent to the requirement that 
the derivative of $\Omega$ with respect 
to the gap parameters vanishes. 
The 
set of gap equations (\ref{sol:gap2}) 
corresponds to five differential equations 
$\partial\Omega/\partial \vec{l}\,|_{\rm min}=0$ where 
$\vec{l}=(b,c,d,e,f)$ at the potential minimum, so we
have analytic expressions for 
$\partial\Omega/\partial \vec{l}$
for any value of the gap parameters $\vec{l}$.
At a first order phase transition,
$\Omega$ should have 
two degenerate minima, one corresponding to the CFL phase and 
the other to the 2SC phase.
We have found two minima at $M_s=235~\MeV$, $\mu=400~\MeV$. (See
Figure 6.)  We define a straight line $\vec l(x)$ in the
five-dimensional space of gap parameters which goes from the 2SC
minimum $\vec{l}(0)=(b=100,c=0,d=0,e=-100,f=0)~\MeV$ to the CFL
minimum $\vec{l}(1)=(91,56,4,-88,-52)~\MeV$.  We evaluate $(\partial
\Omega/ \partial \vec{l}\, )$ along this line, and then obtain the
potential itself by integrating $(\partial \Omega/ \partial \vec{l}\,
) (\partial \vec{l}/\partial x)$ with respect to $x$. The result is
shown in Figure~\ref{fig:firstorder}.  The phase transition is first
order.

The model-independent arguments of Section 2 lead to 
the criterion (\ref{criterion}) for the 
location of the unlocking phase transition. We now
compare this prediction to the quantitative results
we have obtained in our model. For $\mu=400~\MeV$, at the critical
strange quark mass $M_s^{\rm unlock}=235~\MeV$ (see Figure 6) the
gaps $c$ and $-f$, which vanish
in the 2SC phase, are within a few
MeV of each other in the CFL phase, with $c\simeq -f \simeq 55 \MeV$.
Taking $\Delta_{us}\approx 55~\MeV$, we find 
\begin{equation}
\label{results:criterion}
(M_s^{\rm unlock})^2 \approx 2.5 \mu \Delta_{us}\ .
\end{equation}
The unlocking phase transition in Figure 5 occurs at 
$\Ms=350~\MeV$, $\mu=530~\MeV$ and has $c\approx -f\approx 85~\MeV$;
this yields a relation as in (\ref{results:criterion}), 
with 2.5 replaced by 2.7.
We conclude that (\ref{criterion}), derived by simple
physical arguments based on comparing the mismatch between
the up and strange Fermi momenta with the $\Delta_{us}$ 
gap, is a very good guide to
the location of the unlocking phase transition.

\section{Quark-hadron continuity}
\label{sec:cont}

As has been emphasized in Section~\ref{sec:sol}, for low enough
$\ms$ the CFL phase may consist of hadronic matter at low $\mu$, and
quark matter at high $\mu$. This raises the exciting possibility
\cite{SchaeferWilczek} that properties of sufficiently dense 
hadronic matter could be found by extrapolation from the quark matter
regime where models like the one considered in this paper can be used
as a guide at moderate densities, and where the QCD gauge coupling
becomes small at very high densities.

\begin{table}[htb]
\newlength{\wid}\settowidth{\wid}{XXX}
\def\st{\rule[-1.5ex]{0em}{4ex}} 
\begin{tabular}{lccc|cccc} %@{\protect\phantom{XX}}
\hline
\st Quark & $SU(2)_{{\rm color}+V}$ & $Q'$ & gap & 
   Hadron & $SU(2)_{V}$   & $Q$ & gap \\
\hline
\multirow{2}{\wid}{$\left(\ba{c} bu\\[1ex] bd \ea\right)$} & 
\multirow{2}{2em}{\bf 2} &
$+1$ &
\multirow{4}{2em}[-1ex]{$ f$} &
\multirow{2}{4em}{$\left(\ba{c} p\\[1ex] n \ea\right)$} & 
\multirow{2}{2em}{\bf 2} &
$+1$ &
\multirow{4}{2em}[-1ex]{$\De^B_4$} \st \\
& & 0 & & & & 0 \st \\
%\cline{1-3}\cline{5-7}
\multirow{2}{\wid}{$\left(\ba{c} gs\\[1ex] rs \ea\right)$} & 
\multirow{2}{2em}{\bf 2} &
0 & &
\multirow{2}{4em}{$\left(\ba{c} \Xi^0\! \\[1ex] \Xi^-\!\! \ea\right)$} & 
\multirow{2}{2em}{\bf 2} &
0 \st \\
& & $-1$ & & & & $-1$ \st \\
\hline
\multirow{3}{\wid}{$\left(\ba{c} ru-gd\\[1ex] gu\\[1ex] rd \ea\right)$} & 
\multirow{3}{2em}{\bf 3} &
0 & 
\multirow{3}{2em}{$ e$}&
\multirow{3}{4em}{$\left(\ba{c} \Si^0 \\[1ex] \Si^+ \\[1ex] \Si^- \ea\right)$} & 
\multirow{3}{2em}{\bf 3} &
0 &
\multirow{3}{2em}{$\De^B_3$}\st \\
& & $+1$ & & & & $+1$ \st \\
& & $-1$ & & & & $-1$ \st \\
\hline
$ru+gd+\xi_- bs$\hspace{-1em} & \hspace{-2em} {\bf 1} & 0 & $\De_-$ & 
  $\La$ \hspace{-1em} & \hspace{-2em} {\bf 1} & 0 & $\De^B_1$ \st \\
\hline
$ru+gd-\xi_+ bs$ & \hspace{-2em}{\bf 1} & 0 & $\De_+$ &
  --- &  \st \\
\hline
\end{tabular}
\[
\ba{rcr@{}l}
\De_\pm &=& \half  & \Bigl( 2b+e+d \pm \sqrt{(2b+e-d)^2+8c^2}\Bigr) \\
\xi_\pm &=& -{1\over 2c} & \Bigl( 2b+e-d  \mp \sqrt{(2b+e-d)^2+8c^2}\Bigr)
\ea
\]
\caption{Comparison of states and gap parameters in high density quark 
and hadronic matter.}
\label{tab:qm}
\end{table}

The most straightforward application of this idea is to relate the
quark/gluon description of the spectrum to the hadron 
description of the spectrum in the CFL 
phase.\cite{SchaeferWilczek}  
As $\mu$ is decreased from the regime in which
a quark/gluon 
description is convenient to one in which a baryonic
description is convenient, there is no change in symmetry 
so there need be no transition: the spectrum of the theory may
change continuously. Under this mapping, 
the massive gluons in the CFL phase
map to the octet of vector
bosons;\footnote{The singlet vector boson in the hadronic phase
does not correspond to a massive gluon in the CFL phase. This
has been discussed in Ref.~\cite{SchaeferWilczek}.}
the Goldstone bosons associated with chiral symmetry breaking
in the CFL phase 
map to the pions;  
and the quarks map onto baryons. Pairing
occurs at the Fermi surfaces, and we therefore expect the gap
parameters in the
various quark channels, calculated in Section~\ref{sec:res}, to map to
the gap parameters due to baryon pairing.

In Table~\ref{tab:qm}
we show how this works for the fermionic states in 2+1 flavor QCD.
There are nine states in the quark matter phase. We show how they
transform under the unbroken ``isospin'' of $SU(2)_{{\rm color}+V}$ and their
charges under the unbroken ``rotated electromagnetism'' generated
by $Q'$, as described in Section 4. 
Table~\ref{tab:qm} also shows the baryon octet,
and their transformation properties under the symmetries
of isospin and electromagnetism that are unbroken in sufficiently
dense hadronic matter. Clearly there is a correspondence between
the two sets of particles.\footnote{The
one exception is the final isosinglet. 
In the $\mu\to\infty$ limit, where the
full 3-flavor symmetry is restored, it becomes an $SU(3)$ singlet,
so it is not expected to map to any member of the baryon octet.
We discuss this further below.
The gap $\De_+$ in this channel is twice as large as the others
(it corresponds to $\De_1$ in Ref.~\cite{ARW3}).}
As $\mu$ increases, 
the spectrum described in Table 2 may evolve continuously 
even as the language used to describe it changes from baryons, 
$SU(2)_{V}$ and $Q$ to quarks, $SU(2)_{{\rm color}+V}$ and $Q'$.

If the spectrum changes continuously, then in particular so must the
gaps.  As discussed above, and displayed explicitly in
\eqn{sol:ansatz}, the quarks pair into rotationally invariant,
$Q'$-neutral, $SU(2)_{{\rm color}+V}$ singlets.  The two doublets of
Table \ref{tab:qm} pair with each other, with
gap parameter $f$.  The triplet pairs with itself, with gap 
parameter $e$.  Finally, the two singlets pair with themselves.

When we map the quark states onto baryonic states, 
we can predict that the baryonic pairing scheme that will occur
is the one conjectured in Sect.~\ref{sec:int} for sufficiently
dense baryonic matter:
\beq
\ba{ll}
\<p\Xi^-\>,\<\Xi^-p\>,\<n\Xi^0\>,\<\Xi^0n\> 
& \rightarrow \hbox{4 quasiparticles, with gap parameter~~} \De^B_4 \\[0.3ex]
\<\Si^+\Si^-\>,\<\Si^-\Si^+\>,\<\Si^0\Si^0\> 
& \rightarrow\hbox{3 quasiparticles, with gap parameter~~} \De^B_3 \\[0.3ex]
\<\La\La\> & \rightarrow\hbox{1 quasiparticle,\phantom{s} with 
gap parameter~~} \De^B_1
\ea
\eeq
The baryon pairs are rotationally-invariant, $Q$-neutral, $SU(2)_V$
singlets.  It seems reasonable to conclude that as $\mu$ is increased
the baryonic gap parameters $(\De^B_4, \De^B_3, \De^B_1)$ may evolve
continuously to become the quark matter gap parameters $(f, e, \De_-)$, 
which we have calculated in this paper.  Assuming continuity, the magnitude of
the gaps will change as the density is increased but if their ratios
change less we can use our results for the gap parameters in the CFL phase at
$\mu=400~\MeV$ and $M_s=150~\MeV$ to suggest baryonic gap parameters with 
ratios
\begin{equation}
\Delta^B_4:\Delta^B_3:\Delta^B_1\sim 1.06 : 1.26 : 1 
\label{prediction}
\end{equation}
in matter which is sufficiently dense
but still conveniently described as baryonic.

As mentioned above, the ninth quark corresponds to a singlet baryon
which is very heavy, for reasons which have nothing to do with our
considerations. We therefore expect that as $\mu$ is increased
and a quark matter description takes over from a baryonic description,
the density of the singlet quark/baryon becomes nonzero at some $\mu$,
and begins to increase.  Deep into the quark matter phase, there is 
a gap $\Delta_+$ for this ninth quark, but this does not correspond
to any gap deep in the baryonic phase, because the density of the 
corresponding baryon is zero.  There is no change in symmetry
at the $\mu$ at which the density of the ninth quark/baryon
becomes nonzero, just as there was no change in symmetry 
at the lower $\mu$ (within the ``strange baryon'' phase of
Figure 1) at which the $\Xi$ density became nonzero.  Both
these onsets are smooth transitions at arbitrarily small 
but nonzero temperature.  We do not expect the onset
of a nonzero density for the ninth quark/baryon to upset
the continuity between the baryonic and quark matter phases,
but symmetry arguments can only demonstrate 
the possibility of continuity; they cannot prove that there
is no transition.

The analysis leading to (\ref{prediction}) 
provides the first example of the use of the 
hypothesis of quark-hadron continuity in the color-flavor
locked phase to map a quark matter calculation onto
quantitative properties of baryonic matter.

\section{Color-Flavor Locking at Asymptotic Densities}

\begin{figure}[t]
\epsfxsize=4.5in
\begin{center}
\hspace*{0in}
\epsffile{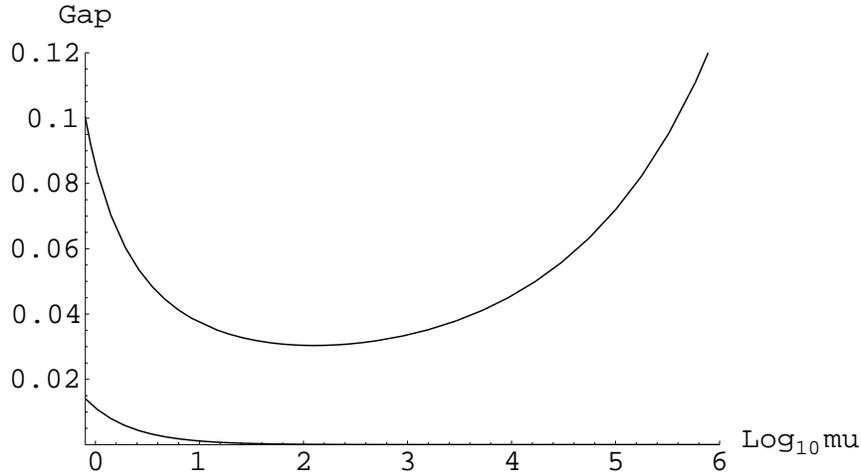}
\end{center}
\caption{The upper curve shows Son's result for the 
superconducting gap $\Delta$ as 
a function of $\log_{10}\mu$, for $\mu$ 
from $0.8$ GeV to $10^6$ GeV. The vertical scale
has been normalized so that $\Delta=0.1$ GeV at $\mu=0.8$ GeV. 
We have taken $g(\mu)$ from the two-loop beta function
for three flavor QCD with $\Lambda_{\rm QCD}=200~\MeV$. Color-flavor
locking occurs whenever $\Delta \gtrsim M_s^2/2\mu$. 
The lower
curve is $M_s^2/2\mu$, taking $M_s=150~\MeV$. 
We conclude that QCD at very high
densities is in the CFL phase.}
\label{fig:son}
\end{figure}
Son \cite{Son} has studied color superconductivity
at much higher densities
than those we can treat with our model, 
using a resummed gluon propagator rather than a 
point-like four-quark interaction, and has determined the 
leading behavior of the gap in the large $\mu$, small $g(\mu)$ 
limit.
Here, $g(q)$ is the QCD gauge coupling at momentum transfer $q$.
The resulting expression,
\begin{equation}
\Delta \sim \mu % \left(\frac{1}{g(\mu)}\right)^5 
\frac{1}{g(\mu)^5}
\exp\Bigl(-\frac{3\pi^2}{\sqrt{2}}
\frac{1}{g(\mu)}\Bigr)\ ,
\label{songapequation}
\end{equation}
describes the $\mu$-dependence of the gap(s), but not
their absolute normalization.
$\Delta$ could be any of our gaps. The distinction between
$b$, $c$, $d$, $e$ and $f$ is in the 
prefactor which should appear in (\ref{songapequation}), but which
has so far not been calculated.  We take our model estimates
as a guide at moderate densities, around $\mu\sim 400-800$ MeV,
and use them to normalize the calculation of Ref.~\cite{Son}
by fixing the prefactor in (\ref{songapequation}) so that  
$\Delta=100~\MeV$
at $\mu=800~\MeV$.  We show the result in Figure \ref{fig:son}.
Note that $\Delta$ is plotted versus $\log\mu$; it changes
very slowly.  It decreases by about a factor of three
as $\mu$ is increased to around 100 GeV (!) and then 
begins to rise without bound at even higher densities.
We conclude that independent of any details (like
the precise value of $M_s$, for example) at asymptotically 
high densities $\Delta$ is 
far above $M_s^2/2\mu$.  For any finite value of the strange
quark mass $m_s$, quark matter is in the color-flavor locked
phase, with broken chiral symmetry, at arbitrarily high
densities where the gauge coupling becomes small.

\section{Conclusions}
\label{sec:concl}

We have discussed
a conjectured phase diagram for 
$2+1$ flavor QCD as a function of $\mu$, the chemical potential for quark
number, and $m_s$, the strange quark mass. We 
have worked at zero temperature
and ignored electromagnetism and the $u$ and $d$
quark masses throughout.
Our quantitative analysis
is restricted to the high density quark matter phases, where we
use a model with an effective four-fermion interaction which has the
index structure of single gluon exchange. 
%However, we present evidence that the
%expected features of 
%superfluid hadronic matter at lower densities match
%the qualitative properties of color superconducting quark matter
%for a small enough or realistic strange quark mass.  

Our analysis shows that for any finite $\ms$ the quarks
will always pair in a color-flavor-locked (CFL) state
at arbitrarily large chemical potential, breaking chiral
symmetry.  What happens at lower $\mu$ depends on the strange quark
mass.  If $\ms$ is greater than a critical value $\ms^{\rm cont}$,
then as $\mu$ is reduced, there will be an unlocking phase transition
to a color superconductor phase analogous to
that in two flavor QCD , with restoration of chiral
symmetry. We give model-independent arguments that
the unlocking transition between the CFL and 2SC phases 
is controlled by the mismatch between the light and strange
quark Fermi momenta and is therfore first order.
We confirm this quantitatively in our model.
If $\mu$ is reduced further, there will presumably be a phase transition to
nuclear matter, and chiral symmetry will be broken again.  However, if
$\ms<\ms^{\rm cont}$ then the 2SC state never occurs. 
The quark matter stays in the color-flavor locked phase
all the way down until the transition to hadronic matter. Chiral
symmetry is never restored at any $\mu$.  In this case there may be
continuity between the CFL quark matter and sufficiently dense
hadronic matter. 
Assuming such continuity, we use our calculation
of quark gaps to make predictions about the gaps in hadronic matter
\eqn{prediction}.

At arbitrarily high densities, where the QCD gauge coupling is small,
quark matter is always in the CFL phase with broken chiral symmetry. This is
true independent of whether the ``transition'' to quark matter is
continuous (as may occur for small $\ms$, including, we surmise,
realistic $\ms$) or whether, as for larger $\ms$, there are two first
order transitions, from nuclear matter to the 2SC phase, and
then to the CFL phase.

There are many directions in which this work can be developed.  We
have worked at zero temperature, so a natural extension would be to
study the effects of finite temperature. The phase diagram of 
two-flavor QCD as a function of baryon density, temperature 
and quark mass has been explored in Ref.~\cite{BergesRajagopal}.
A nonperturbative approach beyond the mean field approximation
that we have employed can be performed along the lines of 
Ref.~\cite{BJW} using the exact renormalization group,
or by doing a lattice calculation \cite{HandsMorrison}. 

Within our model, one could
study more exotic channels that we have ignored, such as
those with $S=1$ and/or $L=1$, or channels that would lead to 
pairing of the strange quarks in 
the 2SC+s phase. 
There are also improvements that could be made to the model,
such as including four- and six-fermion 
interactions induced by the three-flavor
instanton vertex, or
using sum rules to include nonperturbative gluons \cite{AKS}, or
including perturbative gluons \cite{Son}. It would also be desirable
to include the effects of electromagnetism, and of different chemical
potentials for the different flavors.

Finally, it is of great importance to investigate the consequences of
our findings for the phenomenology of neutron/quark stars, which are
the only naturally occuring example of cold matter at the densities we
have studied.  We are confident that, with the basic symmetry
properties of the phase diagram now at hand, a whole new
phenomenology waits to be uncovered as the role of the strange quark
in dense matter is fully elucidated.

\bigskip
\begin{center}
Acknowledgments
\end{center}
We thank E. Shuster and D. T. Son for helpful discussions.  Related
issues are discussed, with a different emphasis, in ``Quark
Description of Hadronic Phases'' by T. Sch\"afer and F. Wilczek, IAS
preprint IASSNS-HEP 99/32.  We thank these authors for showing us
their work prior to publication and for informative discussions.
This work is supported in part  by the U.S. Department
of Energy (D.O.E.) under cooperative research agreement \#DF-FC02-94ER40818.

\appendix

\section{The general form of the gap equation}
\label{sec:appendix}
We calculate the gap equations for a quark-quark condensate
with completely general color-flavor 
structure, using a four-fermion interaction vertex with the
index structure of single-gluon exchange.
We will allow condensation in the channels 
$\De^{\al\be}_{ij}=\<q^\al_i C\ga_5 q^\be_j\>$
and $\ka^{\al\be}_{ij}=\<q^\al_i C\ga_5\ga_4 q^\be_j\>$, with
arbitrary color-flavor structure. We will also allow
the quark mass matrix $\phi^{\al\be}_{ij}$ to have arbitrary
color-flavor structure.
We use the Euclidean conventions of Ref.~\cite{BergesRajagopal}

The mean field gap equation is
\beql{app:gap0}
\Si = -{G\over (2\pi)^4} \int \!d^4q  V_{A\mu} \Minv(q) V_A^\mu,
\eeql
Since we want to study quark-quark condensation, we use the Nambu-Gorkov
basis, in which the quark-gluon vertex has the standard index structure
\beql{app:vertex}
V_{A\mu} = \left(\ba{cc} & -\la^T_A \ga^T_\mu\\ \la_A \ga_\mu &
\ea\right),
\eeql
and the fermion matrix $M$ is
\beql{app:ferm}
\ba{rcl}
M(q) &=& \left(\ba{cc} & C \qplslash C\\ \qmislash \ea \right) + \Si, \\[1ex]
q\!\!\!/_\pm &=& (q_0\pm i\mu)\ga^4 + q_i\ga^i .
\ea
\eeql
$C$ is the Dirac charge conjugation matrix.
In this appendix we allow $\De,\ka,\phi$ to have arbitrary color-flavor
structure, so we treat them as general non-commuting matrices in
\beql{app:self}
\Si = \left(\ba{cc}
C \ga_5(\De + \ka \ga_4) & -i\phi \\ i\phi & (\De + \ka \ga_4)\ga_5 C
\ea\right).
\eeql
We will fix the quark masses $\phi$, and solve for the superconducting 
gaps $\De$ and $\ka$. We can just take the
top left component of \eqn{app:gap0}, which gives us
\beql{app:gap1}
C \ga_5(\De + \ka \ga_4) = {G\over (2\pi)^4} \int \!d^4q
\la^T_A \ga^T_\mu \,X^{-1} \,\la_A \ga^\mu,
\eeql
where $X^{-1}=(\Minv)_{22}$, and $X$ can be written
\beql{app:X}
\ba{rcl}
X &=& \ga_5\Bigl( X_1 I + X_4 \ga^4 + X_V q_i\ga^i 
 + X_S q_i\si^{i4}\Bigr)C \\[1ex]
X_1 &=& (q^2+\mu^2)R + \phi R \phi + i(q_0-i\mu)\De^{-1}\ka R \phi
 -i(q_0+i\mu)\phi \De^{-1}\ka R  + \De \\
X_4 &=& i(q_0-i\mu) R \phi -i(q_0+i\mu) \phi R
 + (q_0^2 - \qvsq + \mu^2)\De^{-1}\ka R
 + \phi \De^{-1} \ka R \phi - \ka \\
X_V &=& 2q_0 \De^{-1} \ka R + i(R\phi - \phi R) \\
X_S &=& 2\mu R + \De^{-1}\ka R \phi + \phi \De^{-1} \ka R \\
R &=& (\De - \ka \De^{-1} \ka)^{-1}.
\ea
\eeql
$X$ can now be inverted,
\beql{app:Xinv}
\ba{rcl}
X^{-1} &=& -C \Bigl( P_1 I + P_4 \ga^4 + P_V q_i\ga^i 
 + P_S q_i\si^{i4}\Bigr)\ga_5 \\[1ex]
P_1 &=& -(U_4^{-1} U_1 - T_4^{-1} T_1)^{-1} U_4^{-1}S_1^{-1} \\
P_4 &=& -T_4^{-1} T_1 P_1 \\
U_1 &=& -S_2^{-1}(-X_4-i \qvsq X_S X_1^{-1} X_V)
        +S_1^{-1}(\qvsq X_V X_1^{-1} X_V - X_1) \\
U_4 &=& -S_2^{-1}(-X_1 + \qvsq X_S X_1^{-1} X_V)
        +S_1^{-1}(i \qvsq X_V X_1^{-1} X_S - X_4) \\
T_1 &=&  S_2^{-1}(-X_4-i \qvsq X_S X_1^{-1} X_V)
        -S_3^{-1}(-X_S + i X_4 X_1^{-1} X_V) \\
T_4 &=&  S_2^{-1}(-X_1 + \qvsq X_S X_1^{-1} X_S)
        -S_3^{-1}(i X_V - X_4 X_1^{-1} X_S) \\
S_1 &=& \qvsq X_S + i q^2 X_V X_1^{-1}X_4 \\
S_2 &=& \qvsq (i X_V + X_S X_1^{-1} X_4) \\
S_3 &=& X_1 - X_4 X_1^{-1} X_4
\ea
\eeql
We have ignored $P_V$ and $P_S$ because these multiply  odd powers of momentum,
and so give no contribution to the integral on the RHS of the gap equation
\eqn{app:gap1}.

Note that in general $P_4\neq 0$, so it is necessary to include
$\ka$ in the ansatz. Without it the gap equations do not close.
That is, if one sets $\ka$ to zero, one finds that if $\Ms$
and $\Delta$ are both nonzero, one obtains $P_4 \neq 0$ and 
concludes that setting $\ka=0$ is inconsistent.
However, $\ka$  turns out to be small, and we neglect it.
To check that $\ka$ is small, we solved the coupled gap equations for
the $\ka$ and $f$ gaps, in one of the $2\times 2$ blocks of
the ansatz \eqn{sol:ansatz}, using a modified gluon vertex so that
this block did not mix with the others. We calculated
$\ka$ and $f$ as a function of $\Ms$, at $\mu=400~\MeV$.
The values of $f$ were very close to those obtained for the full
gluon vertex (see Figure~\ref{fig:gap-mu0.4}), and $\ka$ was zero
at $\Ms=0$, and always less than $6~\MeV$.

% FindfkGap[-30,0.3,0.8,0]    {0.100618, -3.23517 10^(-15)}
% FindfkGap[-30,0.4,0.8,0.05] {0.0996176, -0.00157549}
% FindfkGap[-30,0.4,0.8,0.1]  {0.0965947, -0.0030621}
% FindfkGap[-30,0.4,0.8,0.2]  {0.0842382, -0.00539207}
% FindfkGap[-30,0.4,0.8,0.3]  {0.0630845, -0.00617099}

\end{document}